# Potential application of network descriptions for understanding conformational changes and protonation states of ABC transporters

Tamás Hegedűs[1,2,3,#], Gergely Gyimesi[1,3,#], Merse E. Gáspár[4], Kristóf Z. Szalay[4], Rajeev Gangal[4], and Peter Csermely[4,*]

[1]*Membrane Research Group, Hungarian Academy of Sciences, Budapest, Hungary;* [2]*Department of Molecular Pharmacology, Research Center for Natural Sciences, Hungarian Academy of Sciences, Budapest, Hungary;* [3]*Department of Biophysics, Semmelweis University, Budapest, Hungary;* [4]*Department of Medical Chemistry, Semmelweis University, Budapest, Hungary*

**Abstract:** The ABC (ATP Binding Cassette) transporter protein superfamily comprises a large number of ubiquitous and functionally versatile proteins conserved from archaea to humans. ABC transporters have a key role in many human diseases and also in the development of multidrug resistance in cancer and in parasites. Although a dramatic progress has been achieved in ABC protein studies in the last decades, we are still far from a detailed understanding of their molecular functions. Several aspects of pharmacological ABC transporter targeting also remain unclear. Here we summarize the conformational and protonation changes of ABC transporters and the potential use of this information in pharmacological design. Network related methods, which recently became useful tools to describe protein structure and dynamics, have not been applied to study allosteric coupling in ABC proteins as yet. A detailed description of the strengths and limitations of these methods is given, and their potential use in describing ABC transporter dynamics is outlined. Finally, we highlight possible future aspects of pharmacological utilization of network methods and outline the future trends of this exciting field.

## 1. INTRODUCTION

ABC transporters (ATP Binding Cassette transporters) have been in the forefront of pharmacological studies for many years, since several severe human illnesses, such as cystic fibrosis, Dubin-Johnson syndrome or adrenoleukodystrophy have been linked to mutations leading to the dysfunction of certain ABC transporters [1–4]. In addition, multidrug-resistance of cancer cells has been shown to be mediated by ABC transporters responsible for the extrusion of xenobiotics from cells [5]. Thus, ABC transporters are popular target candidates for the design of drugs to either rescue or attenuate transporter activity.

---

[#]Contributed equally to this work.
[*]Address all correspondence to this author at the Department of Medical Chemistry, Semmelweis University, H-1444 Budapest, P. O. Box 260, Hungary; Tel: +36-1-459-1500; Fax: +36-1-266-3802; E-mail: csermely.peter@med.semmelweis-univ.hu

*ABC exporters* harvest the energy of ATP binding or hydrolysis to actively transport their substrates into the extracellular medium [6]. Mutagenic experiments and structures determined by X-ray crystallography suggest that substrate and ATP binding occur in spatially distant regions of the protein [7–12]. The nucleotide binding domains (NBDs) act in a pair, and bind two ATP molecules cooperatively on their common interface [13], while in known cases drug binding and transport occur in the transmembrane domains (TMDs) [7–9,12]. ATP binding and hydrolysis are necessary for drug transport [6], and the binding and transport of drugs, in turn, stimulate transporter ATPase activity [9,14,15], suggesting the existence of active communication between the drug binding site(s) and the NBDs.

Available X-ray structures of *exporter-type* full transporters (Sav1866 [10], MsbA [11], mMDR1A [12], TM287/288 [unpublished, PDB IDs: 3QF4, 3QF5], hABCB10 [unpublished, PDB IDs: 2YL4, 4AA3]) and the variety of their transported substrates suggest that ABC transporters are very flexible molecules that can achieve various conformations in the membrane during the transport cycle. Several mechanistic models have been developed to describe the details and the order of the putative events in the transport cycle [6,13,16–19]. However, a consensus mechanism of action has not been obtained that could be applicable to these ABC transporters. To understand the mechanism how drugs stimulate the ATPase activity, to elucidate the action of inhibitors, and the mode of transport of large substrates, the consideration of the dynamic behavior of ABC transporters is necessary.

In the last decades network related methods emerged as powerful tools to determine key properties of protein structure and dynamics. In the representations of protein structure networks (also called protein contact networks, or residue interaction networks) amino acid side chains represent the nodes of the network. The edges of protein structure networks are derived from the distance between the side chains [20–27]. Other network-related methods, such as elastic network models, the statistical mechanics approaches of Gaussian network models and normal mode analysis, proved to be extremely useful to explore protein dynamics [28–37].

In this review we describe the strengths and limitations of these network representations in detail, and will show their usefulness in the description of conformational and protonation changes. First we summarize key aspects of conformational and protonation changes of ABC transporters and their importance in pharmacological design. In Section 3 several potent, network-related tools to assess conformational and protonation changes are described. We highlight several potential applications of the dynamic network description of ABC transporters in Section 4 and conclude the review by listing the most important trends of the field.

## 2. ROLE OF CONFORMATIONAL CHANGES AND PROTONATION STATES IN THE FUNCTION OF ABC TRANSPORTERS

**2.A. Conformational transitions of ABC transporters**. The structural elucidation of ABC transporters has not started until recently because the crystallization of large membrane proteins is tedious and challenging. Still, several atomic resolution structures of both bacterial and eukaryotic full ABC transporters have been solved (BtuCD [38], HI1470/71 [39], ModBC [40], MalFGK$_2$ [41], MetNI [42], Sav1866 [10], MsbA [11], mMDR1A [12], PDB IDs without reference: hABCB10 [2YL4], TM287/288 [3QF4]). While all of them share a similar domain composition having two nucleotide binding (NBD) and two transmembrane domains (TMD) per active transporter, the number of domains on one polypeptide chain and the topological organization of transmembrane helices vary.

Among the ABC proteins with available crystal structures, bacterial *ABC importers* feature all four domains on separate chains and contain either 10+10 transmembrane helices forming a densely packed structure (structures of BtuCD, HI1470/71) or 6+6 transmembrane helices (structures of ModBC, MetNI, MalFGK$_2$). Bacterial and eukaryotic *ABC exporters* with available structure usually contain a TMD and an NBD on one polypeptide chain (structures of Sav1866, MsbA, TM287/288, hABCB10) and function as homo- or heterodimers, or have all four domains on a single chain (mMDR1A). Exporters also feature 6+6 transmembrane helices, although organized in a different topology than importers. In spite of their similar domain composition, the different membrane topologies suggest that the mechanism of action may not be similar in all ABC transporters [43].

Conformations of *ABC exporters* seen in currently available X-ray structures can roughly be classified into three groups, based on the separation between the NBDs and the orientation of the transmembrane domain (Fig. **1**). In the presence of nucleotides, the transporters form a so-called "bottom-closed" conformation, where the two NBDs associate to form a tightly bound dimer-like structure, and the transmembrane domains are in an outward-facing conformation (PDB IDs: 2HYD, 3B60). These structures also reveal that ATP binding occurs at two pseudo-symmetric sites on the interface of the associated NBDs, where the nucleotides are sandwiched between the two protein domains. In crystals without nucleotides (the apo state), the transmembrane domains are in the inward-facing conformation, and the nucleotide binding domains do not associate. The extent of separation between the NBDs in the apo structures show large variations ranging from 30 to 70 Å, which further divides the inward-facing conformations into a group where the NBDs are in contact with each other (PDB IDs: 3B5X, 2YL4), and to another group where the NBDs are completely separated (PDB IDs: 3B5W, 3G5U). It is remarkable that a single transporter can display conformations belonging to either group, as demonstrated by the three independent structures of the MsbA exporter (PDB IDs: 3B60, 3B5X, 3B5W). These three conformational groups are often viewed as representing different states along the transport cycle. These structures also fulfill the first two of the three basic requirements of the simple membrane pump model proposed by Jardetzky [44], by containing a cavity that can open toward either side of the membrane, and is large enough to accept substrates. The model requires that substrate binding sites reside in the cavity, which is also suggested by mutagenesis studies [7,8,45] and by the mouse MDR1A X-ray structure with a bound inhibitor [12]. The substrate binding cavity should also have lower affinity in the outward-facing state than in the inward-facing state to facilitate the dissociation of the transported substrate. It has been suggested based on structural and drug binding data that rotation of transmembrane helices during the conformational switch to the outward-facing state causes crucial drug binding residues to become buried within the domain or even facing the lipid bilayer, which enables the dissociation of the substrate [46].

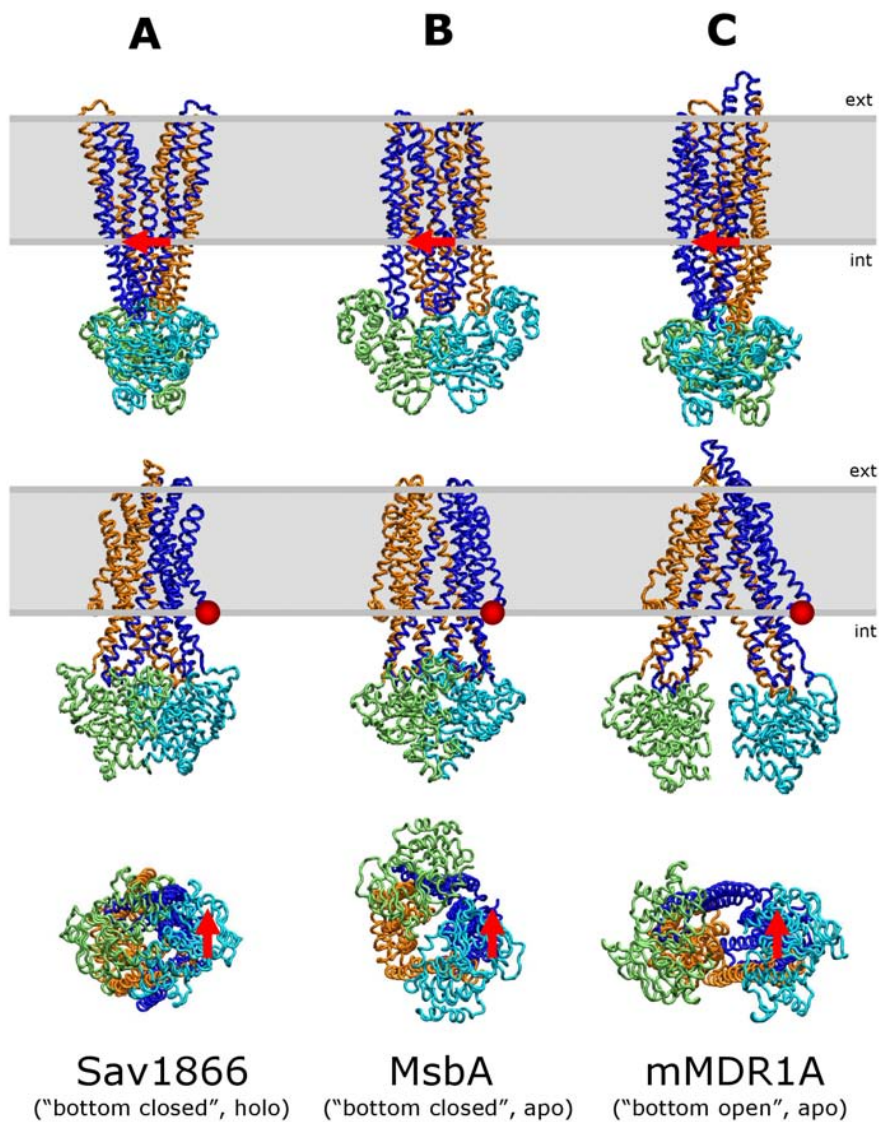

**Fig. (1). Conformations of ABC exporters.** Representatives of the three groups of currently available ABC exporter X-ray structures are shown. (**A**) The nucleotide bound (holo) state is represented by the Sav1866 bacterial exporter (PDB ID: 2HYD), where the nucleotide binding domains are tightly associated and the transmembrane domain is in the outward-facing conformation. (**B**) A "bottom closed" apo conformation is structurally intermediate between the (**A**) and (**C**) states. The NBDs are separated but in contact with each other, and the substrate binding cavity within the transmembrane region is open toward the interior of the cell. This conformation can be observed for the MsbA transporter (PDB ID: 3B5X). (**C**) In the "bottom open" apo conformation, the NBDs are widely separated not contacting each other and the transmembrane domain is clearly in the inward-facing state. This is represented by the mouse MDR1A structure (PDB ID: 3G5U). The red arrows indicate the orientation of the N-terminal "elbow" helices that are parallel to the membrane surface as a common frame of reference. The membrane is shown in grey. (The color version of the figure is available in the electronic copy of the article).

Thus, based on the structures, it is relatively easy to deduce a simple mechanistic model for the transporters. According to this simple model, the switch from the inward-facing to the outward-facing state of the transmembrane domain occurs with the association of NBDs upon nucleotide binding, the substrate binding sites residing in the transmembrane domains. The coupling between the cytoplasmic and the transmembrane regions is provided by short helix segments connecting the cytoplasmic ends of the transmembrane helices. These regions of the transmembrane domains are referred to as *coupling helices* because they are tightly associated with the NBDs in all known structures and are proposed to form a junction that is responsible for allosteric communication between the TMDs and the NBDs [10]. Even such a simple mechanistic model shows that ABC exporters can undergo large conformational changes during the transport cycle.

It should be emphasized that X-ray structures only represent snapshots of the transporter that has been stabilized under the crystallization conditions, and should therefore be considered with caution and scrutinized for consistency with other biochemical data [43]. Several other methods in tandem with structure determination have been used to develop more elaborate mechanisms of *ABC exporter* function. Early models were based on ATPase inhibition studies by vanadate or beryllium-fluoride, which can trap the system in the transition state of hydrolysis [19]. Vanadate trapping experiments in human P-glycoprotein showed that hydrolysis at a single site is enough to drive the transport cycle and suggested an alternating ATP hydrolysis model [13,47]. However, studies on isolated NBD dimers were unable to trap an asymmetric ATP/ADP bound state that would be predicted by the alternating sites model [17]. In addition, it was shown that NBD dimer formation is dependent on ATP binding [17]. In view of these findings, and also supported by X-ray structures of NBD dimers which contained nucleotides in both binding sites, the processive clamp model (or switch model) was proposed [6,17]. This model, in contrast to the alternating sites model, postulates that hydrolysis of both ATPs is required for the dissociation of the NBDs, which is in turn necessary for the commencement of the transport cycle.

The presence of TMDs also seems to affect NBD association and ATP hydrolysis, as substrate binding stimulates ATPase activity only in transporters having both NBD and TMD domains [17,48]. Moreover, several heterodimeric and single-chain full transporters contain a degenerate ATP-binding site, rendering the ATPase activity of that site ineffective [49-51]. Thus, it is not clear how the switch model should be interpreted in these cases. There is also a discrepancy between mechanistic models regarding whether a large and complete separation between the NBDs occurs during transport. While chemical cross-linking studies of Loo and Clarke [52] showed that human MDR1 is able to function while the cytoplasmatic parts of the transmembrane domains are cross-linked, the solved apo structures of the MsbA and mouse MDR1A transporters display a large separation between the NBDs. As of yet, no consensus model of ABC exporter mechanism has been established that is able to reconcile all experimental findings and structures.

Since conformational transitions are hard to probe experimentally, molecular simulation methods have also been used to elucidate conformational changes at important points of the transport cycle. As ATP binding has been shown to promote the association of NBDs, it has been argued that ATP binding rather than hydrolysis can be the driving force behind the transport cycle [53]. Since the ATP-stabilized NBD dimer is very stable in hydrolysis-deficient mutants [17], the hydrolysis of ATP in a fully functional transporter is expected to trigger later events of the transport cycle, such as resetting the original high-affinity drug binding conformation.

Changes in conformation and dynamics occurring directly after ATP hydrolysis are often studied by molecular dynamics simulation, by the *in silico* replacement of ATP with ADP or ADP+$P_i$. These simulations have shown an opening of the NBD interface at the hydrolysis site due to the loss of interactions with the gamma phosphate [19,54,55]. In addition, the so-called helical subdomain of the NBD, nearest to the hydrolyzing site, undergoes an outward rotation similar to the one observed in X-ray structures of monomeric NBDs [19,54-57]. The effect of ATP hydrolysis has also been studied in the full length Sav1866 bacterial exporter either by ADP or ADP+$P_i$ replacement [58-60] or by nucleotide removal [58,60-62]. It was found that the outward-facing cavity of the transmembrane region can undergo spontaneous closure upon hydrolysis [59-61], and ATPase site opening was also observed in these studies [58,59,62], although not to the same extent as in isolated NBD dimers. Some studies have aimed to describe a putative sequence of events in the full transporter, occurring after ATP hydrolysis, and these simulation results suggest that destabilization of the helix bundle, formed by the cytoplasmic extensions of the transmembrane helices, might be one of the first steps in the switch toward the inward-facing conformation [59,63]. The disruption of the bundle opens the inner cavity toward the cytoplasm, and the rearrangement of the transmembrane helices completes the conformation change [63]. While the effect of ATP hydrolysis on the NBDs and TMDs has been extensively studied, the mechanism of ATPase activity stimulation by drug binding, where intramolecular signaling is expected to occur in the reverse direction from the TMDs towards the NBDs, still remains elusive and its structural basis remains to be understood.

ABC transporters are integral membrane proteins, still, in many cases the effects of different lipid environments (e.g. ER versus PM, mammalian versus insect membrane) on the conformation, dynamics and function is neglected. In the case of ABCG proteins the protein function is strongly dependent on the types of sterols specific to the species. The yeast Pdr12p transporter functions in the presence of ergosterol but not in cholesterol [64]. In contrast, the human ABCG2 is active in the presence of cholesterol, which cannot be fully substituted by ergosterol or sitosterol [65]. Nonetheless the latter sterols also form membrane rafts. Interestingly, the Gly and Thr variants of the Arg482 amino acid located presumably in the interface with the inner leaflet of the lipid bilayer have different cholesterol sensitivity that the wild-type ABCG2. All these data suggest that the interaction of ABCG proteins with sterols is highly specific and the sterols do not simply provide an environment for the conformational changes. These lipids may be structural parts of the proteins and serve as cofactors. Cholesterol dependent function of MDR1 has also been reported [66], although this could not be observed with either MDR1 or MRP1 under the same conditions applied for ABCG2 [65].

The role of the above described conformational changes in ABC proteins is to move substrates through a membrane bilayer. Although one large binding pocket exists in currently accepted models, functionally at least two different binding sites are predicted that may be either at different locations of the biding pocket or may physically overlap [14]. These sites exhibit preferential binding of different substrates, e.g. in MDR1 one site binds verapamil, while the other Hoechst 33342. It also seems evident that the induced fit mechanism can explain the wide substrate specificity of MDR transporters. Binding of different substrates has been demonstrated to result in different conformations of the transmembrane helices based on MDR1 cross-linking experiments [67]. However, the conformational selection theory, which is an alternative mechanism for molecular recognition, came to the focus of research again in the last decade [68-72]. According to conformational selection theory, all possible conformations of the protein are realized in the absence of substrates, even those conformations that are capable to bind substrates. The substrate selects the favored conformation to bind to.

In the case of ABC transporters this mechanism would not simply be an alternative of induced fit for recognition of an extremely wide set of drugs, but could also provide explanation for the following phenomena. Multidrug transporters express high basal ATPase activity when the ATP consumption is not coupled to the presence and movement of any drug [73]. Although basal activity is being proposed as a requirement for low substrate selectivity [74], we are far from understanding this phenomenon. According to conformational selection theory, drug binding conformations that are also responsible for ATP cleavage in basal activity are also realized in the absence of any drugs. The presence of a drug would shift the conformational ensemble towards the binding competent conformation exhibiting the increased, drug-stimulated ATPase activity. Unexpectedly, some transported substrates do not increase ATPase activity in the case of several transporters such as yeast Pdr5, Cdr1, and human ABCG2 [49,75,76]. This behavior can be explained by the ATPase competent conformational ensemble being narrowed by the presence of the substrate. The conformation selection model can also provide an explanation of altered substrate specificities when a mutation is located clearly outside a substrate binding region or in the nucleotide biding domains [49,77,78]. Interestingly, the type of the nucleotide has also been reported to alter substrate specificity in Pdr5 [49,79]. In these cases, either the mutation or the modified nucleotide alters the dynamics of the NBD, which is strongly coupled to the transmembrane domains (see CFTR as a well-studied example in Section 2.C), and thus the conformational ensemble of the transporter is changed to favor conformations binding different drugs than the wild type transporter. Interestingly and importantly, thermodynamic analysis indicated that etoposid and colchicine are transported by the G185V variant of MDR1 more efficiently compared to the wild type, since this mutant requires fewer bond rearrangements to reach the transition state [80]. This mutation may shift the structural ensemble to favor conformations that are closer to the transition state of etoposid and colchicine transport. The kinetic substrate selection model, in many ways similar to conformational selection, has been described for ABC transporters [9].

A mechanism similar to conformation selection may be responsible for channel gating of CFTR, where the role of the substrate in changing the structural ensemble is substituted by PKA phosphorylation of the intrinsically disordered R domain of this channel. It has been implied that the phosphorylation of numerous PKA consensus sites leads to altered ordered/disordered states of the R domain, resulting in altered interactions of this regulatory domain with NBD1 [81]. However, if the unphosphorylated R domain intercalates between NBDs, and inhibits transporter function by counteracting NBD dimer formation, then alternative activation mechanisms provided e.g. by dATP [82] or vitamin C [83] cannot be explained. Conceivably, the R domain acts as a constraint between the two halves of the channel, and its phosphorylation results in alterations in the dynamics of CFTR, shifting the conformational ensemble toward a state, which is more likely to open [84]. dATP and vitamin C molecules may act similarly on the unphosphorylated protein by altering the conformational space available.

In order to rationally develop successful pharmacological applications it would be important to learn more about the dynamics of ABC transporters. However, the experimental study of the dynamics of ABC transporters is extremely challenging because these are large membrane proteins difficult to express at sufficient quantities for biochemical studies, and are also sensitive to purification and reconstitution. The methods that can be applied to full length transporters, such as fluorescent quenching and accessibility measurements, have their limitation and yield no directional and quantitative information on protein movements. Although chemical cross-linking experiments provide distance information, they possess a different type of limitation. Since cross-linking is an irreversible process, conformations that are present only with a very low probability can also be detected. Most likely because of this property of cross-linking, a single, most favorable structure determined by crystallization cannot be expected to satisfy all of the measured distances (for a review see [43]). While a distance distribution can be determined between two spin-labeled amino acids using EPR spectroscopy, it might not represent one single conformation, since the distance of two residues can be the same in different conformations. Directional information on the movements is also lacking in EPR experiments. Current computational tools and resources also pose limitations in assessing the dynamics of ABC transporters. For example, molecular dynamics simulations have limited time scale because of the large size of the system (a large protein and lipids to form a bilayer). However, combining of data from different methods, such as NMR spectroscopy of isolated NBDs, cross-linking experiments of the full length proteins and simulation with simplified models may provide solid and sufficient information on dynamics for pharmacological applications.

### 2.B. Protonation States of ABC Transporters

There is very little information about the protonation states of ABC transporters. Their function in most cases does not seem to involve transport of protons, and ABC protein function is not altered by small changes in pH. Interestingly, LmrA, a *Lactococcus lactis* homolog of human MDR1, has been shown to be a symporter of proton and ethidium [85]. Although human MDR1 has also been suggested to function as a proton pump [86], measurable proton transport activity could not be detected [87].

Since it is not trivial to determine the hydrophobicity of amino acid side chains, it is also difficult to learn about their partitioning and $pK_a$ in a membrane environment [88,89]. Recent computational works have explored the protonation states of arginine in a membrane bilayer and found that lipid composition affects the stability of different protonation states [90]. However, the situation is highly complex in a membrane protein, since in many cases the side chain of an amino acid is not exposed to the hydrophobic core of the bilayer. The side chain may point to the interface region or toward another transmembrane helix of the protein and it is difficult to judge if the properties of the environment resemble a hydrophilic or a hydrophobic surrounding, promoting protonation or deprotonation. Arginines located in transmembrane helix 6 have been shown to line the channel pore in CFTR [91,92]. It is proposed that rotational movements make the positively charged arginines inaccessible for chloride ions [93]. This rotation, during which the protonation state of the arginines may or may not change, provides a possible mechanism for fast opening and closing of the channel through two conformational states with relatively low energy barriers. Here we refer to the fast open/close burst, not to the long closed state of the channel upon ATP hydrolysis at the second NBD [94]. Similar rotational movements may be responsible for disrupting and take away the binding sites in multidrug transporters to allow the substrate dissociate from the protein.

The protonation states of histidines in ABC multidrug transporters have turned out to be structurally very important. Recent computational results indicate that the correct protonation state of histidines (the catalytic His583 and His1228, as well as His149 from intracellular loop 1, for the position of the catalytic histidines, see Fig. **2**) and the presence of magnesium ions during simulations with the inward-facing, "bottom open" structure of mouse MDR1A are essential for running stable stimulations (M. L. O'Mara, personal communication). These minor but crucial environmental changes propose a nucleotide free structure, in which the transmembrane domains are separated at the bottom as in the crystal structure, but the nucleotide binding domains get close to each other and form contacts similar to the open state of MalK [95]. In addition to histidines, there are several other residues whose protonation states might be important during ATP hydrolysis, including a glutamic acid that is proposed to form a "catalytic dyad" together with a histidine [96] (Fig. **2**).

In addition to the protonation state of the membrane protein, the protonation state of transported substrate is also important. First, protonation can influence the membrane permeability of drugs, as molecules with charges cannot easily penetrate the bilayer. Second, the protonation state defines whether or not a molecule can accept or donate hydrogen bonds that can be crucial in drug recognition by ABC transporters. Omote and Al-Shawi have studied the effect of protonation state on the membrane permeation of different MDR1 substrates [97]. For example, the deprotonated form of verapamil could enter the hydrophobic core of the bilayer, while the protonated form, similarly to other modeled substrates, maintained its interaction with the head groups of lipids and with water in the interfacial region. Thus the protonation state can determine the most probable position of a certain molecule in the bilayer and so may influence whether that molecule has access to the main drug binding chamber. Since the MDR1 multidrug transporter favors positively charged molecules as substrates [98,99], the entry point of the substrate in the protein may be defined by negatively charged amino acids. The lateral opening between transmembrane helices 4 and 6 could potentially serve as an entry point for substrates in the "bottom open" structure of mouse MDR1A, and the equivalent region in the human MDR1 contains negatively charged residues (e.g. E353 and D997, see Fig. **2**) that may act as selection filters. This presumption is reinforced by the fact that these amino acids are homologous to the E314 of the above mentioned LmrA, in which glutamic acid plays a role in proton transport [100]. The binding chamber contains not only hydrophobic but also charged residues and hydrogen bond acceptors that may form a H-bonding ladder, a pathway for the protonated and positively charged drugs leading toward the extracellular region of the transporter.

### 2.C. Pharmacological Applications of ABC Transporter Dynamics

ABC transporters are important drug targets in many human diseases. In several pathological situations the functional expression of ABC proteins needs to be corrected. The most prominent example is cystic fibrosis, where the mutant forms of CFTR (ABCC7) need to be rescued from a misfolded state [101]. Other diseases in which some of the mutations likely cause abnormal protein folding and decreased function include Stargardt's disease and retinitis pigmentosa (ABCA4; [102,103]), pseudoxanthoma elasticum (ABCC6; [104]). In the case of the treatment of various diseases the function of multidrug transporter ABC proteins should to be inhibited (for recent reviews see [105] and [106]). For example, blocking the function of human MDR proteins in the gut would enhance the absorption of orally administered drugs, in the blood-brain barrier would increase the concentration of drugs in the brain that target the central nervous system (e.g. antidepressants), and in cancer cells would allow reaching the effective cellular concentration of anticancer agents. To develop a successful pharmacological approach that modulates the function of ABC transporters, targeting a static structure is unlikely to be sufficient. Without attempting to be comprehensive, we demonstrate how the knowledge of ABC protein dynamics can be important for pharmacological applications. As an example we employ CFTR, whose dynamics is the most well studied among human ABC transporters.

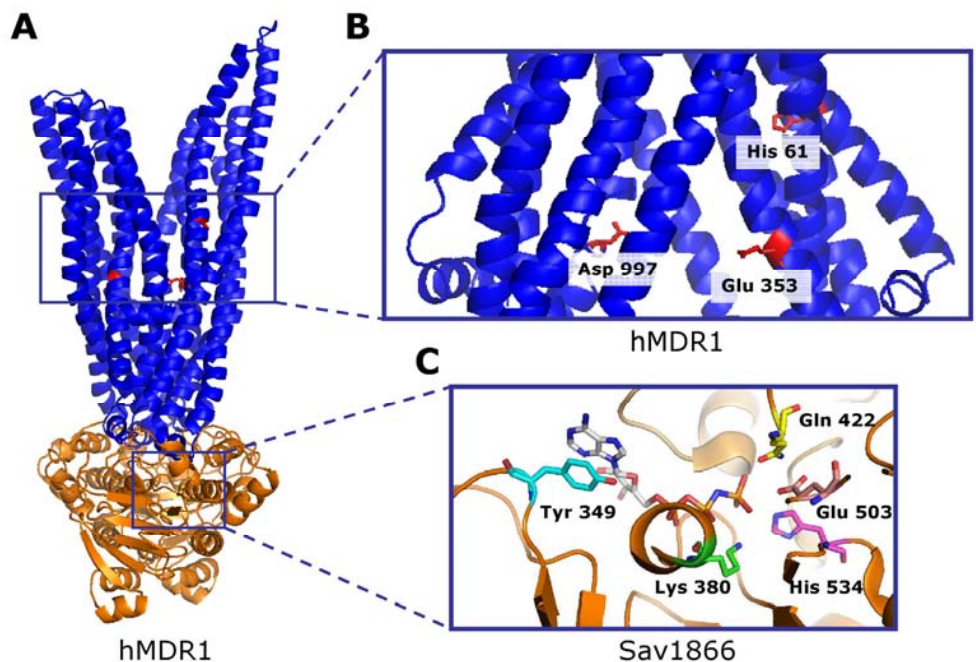

**Fig. (2). Protonation states of ABC transporters.** The protonation state of residues is especially important near the nucleotide binding region, where ATP hydrolysis occurs, and near the putative drug binding regions. The homology models of human MDR1 (**A**) in the outward-facing conformation based on the Sav1866 structure [238] and (**B**) in the inward-facing conformation based on the mouse MDR1A structure are shown as an illustration. Residues whose protonation state could be important in drug binding and selectivity are indicated in red (Glu353, Asp997, His61). (**C**) The ATP binding site is enlarged and canonical residues are highlighted in the Sav1866 AMP-PNP bound structure (PDB ID: 2ONJ) that are important in ATP binding and hydrolysis. While the Tyr349 residue stabilizes the adenosine moiety, several residues, such as Gln422, Glu503 or Lys380 take part in coordinating either the phosphate groups or the magnesium ion of bound MgATP. The His534 residue, which is expected to be doubly protonated [239] is thought to form a "catalytic dyad" with Glu503 [96] and coordinate the water molecule that performs the nucleophilic attack on the phosphate group. (The color version of the figure is available in the electronic copy of the article).

Although many folding mutants of CFTR are known, the deletion of Phe at position 508 (ΔF508) is present in most of the cystic fibrosis patients, thus this mutation is widely studied [101]. Earlier studies based on the Trp fluorescence of isolated polypeptides showed that ΔF508 diverts the CFTR NBD1 domain, where this amino acid is localized, to an off-folding pathway [107,108]. Folding simulations employing a simplified protein model also demonstrated that this deletion alters the folding kinetics of this domain [109]. Although the latter study pointed to specific conformations and specific amino acids that could be targeted to correct the folding pathway, because of the complexity of this process and our limited knowledge on protein folding, drug development and treatment could not be designed based on this observation. It has also been shown that the deleterious effect in the NBD influences domain-domain interactions and also propagates to the transmembrane domains [110,111], since the F508 is a crucial residue at the NBD and TMD interaction interface [112].

The ΔF508 mutation was expected to cause dramatic misfolding of the NBD1, while the X-ray structures of the wild type and mutant isolated domains are highly similar [113]. It is important to note that most of these structures contain mutations that increase solubility or rescue the full length ΔF508 protein. Moreover, the crystallization conditions (e.g. low temperature, high salt concentration) may promote the correct folding and the crystal formation of the mutant polypeptide even if it has decreased thermostability.

In recent years it has become evident that ΔF508 causes alterations in NBD1 dynamics. First, local alterations were demonstrated experimentally in the region where the F508 is localized, employing backbone amide H+/D+ exchange measured by mass spectrometry [114]. These measurements revealed that ΔF508 increases the backbone dynamics at residues 509-511 and in parallel, crystal structures showed surface exposure of V510 in the mutant, which is buried in the wild type protein. Close to this region, the linker between the α- and β-subdomains (residues 492-499) showed increased mobility in the mutant compared to the wild type protein in a molecular dynamics study [115]. This enhanced dynamics also resulted in an increase in exposed surface that may promote the recognition and degradation of the mutant protein by the quality control system in the cellular environment. Interestingly, the removal of the regulatory insertion (RI; residues 404-435), which is present exclusively in the CFTR NBD1 and is highly dynamic [81,113,114], increased the functional expression of ΔF508 CFTR [116]. Although the role of this region is unknown and its removal does not alter channel function in oocytes [117], the deletion of the RI increases the cell surface protein level and the half-life of the mutant ΔF508 protein. Moreover, RI removal also restores domain-domain interfaces, nucleotide binding in NBD1, and channel function to the wild type level. Since the RI region is dynamic and its deletion increases the mobility of the F508 region, it was proposed that these two regions are dynamically coupled. However, analysis of discrete molecular dynamics simulations indicates that this is not the case [116]. In contrast, movements of the F508 region correlate with the dynamics of the β-subdomain. This correlation is lost in ΔF508 and can be partially restored by RI deletion. These results suggest that some of the RI conformations interfere with the dynamic coupling between the F508 loop and the β-subdomain. A potential pharmacological approach could be to develop a small molecule to restrict the dynamics of RI that may restore the coupling between these regions and thus rescue the folding and expression of the mutant protein. Indeed, the intrinsic large structural fluctuations of NBD1 may be the origin of the enhanced effects of ΔF508 in humans compared to other species, where several flexible regions contain Pro residues restricting the available conformational space [118].

Since alterations in dynamics are responsible for the impaired processing and function of the mutant CFTR, the so-called rescue mutations, which are introduced at other locations of NBD1 (at I539 [119], F494/Q637 and F429/F494/Q637 [120] and R553 [121]) and partially restore the functional expression of the ΔF508, act most likely by restoring the dynamic coupling between different parts of ΔF508 NBD1 to the wild type level. These rescue mutations and the observation that the binding of ATP promotes the vectorial folding of CFTR NBD1 [122] show the validity of designing small molecules that interact with the intracellular NBD and restore the folding pathway. ATP binding most likely shifts the dynamic ensemble of nascent chains at a critical step of the co-translational folding promoting the availability of a conformation critical for the further steps of correct folding. Similarly, a water soluble substrate (intracellular glutathione) interacting with the cytoplasmic parts of MRP1 rescues the folding deficient ΔF728 MRP1, which is analogous to the ΔF508 CFTR [123]. Interestingly, affecting the available conformational space in a part most distant than the nucleotide binding domains also seems to be a valid approach to rescue the mutant channel. The ΔY490 MDR1, analogous to ΔF508 CFTR, can be rescued by the interaction of a hydrophobic drug, cyclosporine A [124]. Chemical cross-linking experiments confirmed that cyclosporine A binds to the TMDs promoting the native TM helix conformation. Moreover, at the same time the TMD/NBD interface and the native conformation of NBD are also restored, indicating a strong dynamic coupling between the transmembrane domains and cytosolic parts [124].

In chronic inflammatory bile duct diseases the expression and function of hepatobiliary ABC transporters requires stimulation [125] that could be achieved based on similar principles as described above for CFTR. Mutations that alter the substrate specificity of multidrug transporters expressed at physiological barriers (e.g. intestine, blood-brain barrier) can seriously alter the pharmacokinetics (such as absorption, secretion) of medicaments. This usually leads to increased accumulation of xenobiotics in the body, also increasing the side effects of the particular treatment [126]. If the substrate specificity is defined by dynamics as described in Section 2.A, a small molecule corrector could shift the dynamic ensemble back to wild type and allow treatment with regular doses that were developed and tested in patents carrying wild type alleles.

This complexity of dynamic coupling in ABC transporters has major implications for drug development. Allosteric modulators of dynamics would allow the application of small molecules acting in the membrane region to correct the folding of an intracellular domain. The above examples also show that altered dynamics of a certain region is not fully responsible for the malfunction of a protein. Moreover, it has been shown that ΔF508 alters NBD1 both kinetically and thermodynamically and correction any of these defects results only in partial rescue of the mutant channel [127]. Therefore non-conventional multi-target drugs may be developed to allow the functional rescue of ABC proteins. This type of drugs should target different regions in the protein, thus even actions with partial efficiency may result in full restoration of protein function based on network theory [128-130]. These strategies can also be applied to other ABC transporters with diminished folding and altered dynamics, such as ABCA4 in macular degeneration [102,103] or ABCC6 in pseudoxanthoma elasticum [104].

## 3. NETWORK DESCRIPTIONS OF CONFORMATIONAL AND PROTONATION CHANGES

To understand and modulate the dynamical coupling between regions of proteins, novel theories and tools of networks can be applied. In this Section first we give a brief summary of the definitions of networks related to protein structure and dynamics, such as protein structure networks, protein sectors, Gaussian network models, elastic network models, normal-mode analysis, networks based on the correlations of residue fluctuations and conformational networks. We highlight the use of these network models to predict hot-spots involved in ligand or protein binding, as well as in the transmission of conformational changes. Finally, the applications of the network approach in the assessment of protonation changes and drug design are summarized.

### 3.A. Network Descriptions of Protein Structure and Dynamics

Protein structure networks (which are recently more and more referred to as residue interaction networks, RIN-s) represent the molecular basis of more sophisticated cellular networks, such as protein-protein interaction or signaling networks. In most protein structure network representations not the individual atoms are the nodes of the network, but due to the concerted movements of amino acid side-chains, amino acids of the protein represent network nodes. Network edges are related to the physical distance between side-chains, in most cases measured between $\alpha C$ or $\beta C$ atoms. In some representations the centers of mass of the side chains are calculated, and distances are measured between them. Edges of un-weighted protein structure networks connect amino acids having a distance below a cut-off. The cut-off distance is usually between 4 to 8.5 Å [20-23,25-27,131-134]. Recently a detailed study was published comparing the effect of various $\alpha C$-$\alpha C$ contact assessments, such as the atom distance criteria, the isotropic sphere chain and the anisotropic ellipsoid side-chain models, as well as of the selection of various cut-off distances. The atom distance criteria model proved to be the best, and had a moderate computational cost. The best amino acid pair specific cut-off distances varied between 3.9 and 6.5 Å [135]. Protein structure networks often have weighted links instead of distance cut-offs, where the edge weight is usually inversely proportional to the distance between the two amino acid side-chains. Covalent bonds may be included or excluded in the network representation [20-23,25-27,131-134].

Protein structure networks are "small worlds" meaning that their nodes are separated by only a few edges on average, and the network diameter grows only logarithmically with increasing number of nodes, i.e. with the size of the protein. This small-worldness is very important in the fast transmission of conformational changes, since it means that all amino acids can reach each other by taking only a very little number of steps. The degree distribution of protein structure networks is usually not scale free. This means that we may find much less large hubs (amino acids surrounded by a large number of neighbors) than in other networks. However, the existing smaller hubs play an important role in protein structures, e.g. these 'micro-hubs' were shown to increase the thermodynamic stability of proteins [22,25,28,131,136-141].

The absence of large hubs in protein structure networks is primarily due to the limited changes of protein structures. These changes are limited in the sense that amino acids may not just leave their original neighbors, and associate with another neighborhood, since the covalent bonds do not allow such a gross reorganization. Large changes of network environment are rather usual in networks at higher hierarchical levels. Proteins are often transported from one cellular compartment to the other, where they bind to a different set of other proteins. Several proteins, such as "date hubs" [142] are known for notoriously promiscuous behavior, i.e. for changing their binding partners rather often. In this way to the same binding site a large number of different neighbors may bind at different time points, which - at larger time windows, or observing larger popula-tions, as high-throughput experiments usually do - appears as the existence of large hubs having a large number of neighbors. As another example, several hundreds of acquaintances of the Reader on Facebook do not mean that the Reader is able to conduct a parallel information exchange with all of them at the same time (if doing that right now the Reader probably would not be reading this paper). In agreement with the above considerations, recent investigations reported large changes in the structure of mobile telephone call networks, when the time window of call aggregation was changed [143]. In summary, being a hub is a relative measure, which often depends on the time window of observation. The time window dependence is prevalent in networks where the node neighborhood may be changed, like in social networks. Thus the 'mega-hubs' of social networks are sequential hubs, unlike the 'micro-hubs' of protein structure networks, which are parallel hubs.

Protein structure networks have modules (or in other words, which are mostly used in social sciences: network communities). A network module describes a group of nodes, which are connected (or in a recently used definition: exchange information) more densely with their neighborhoods than the entire group. Modules of protein structure networks quite nicely locate protein domains [26, 27, 139,144-150].

Recently protein structure networks have been used to predict key functional residues, hot-spots, substrate and protein binding areas as high centrality regions (hubs, closeness and betweenness centralities) but having a lower clustering coefficient [151,152]. Some of these positions nicely correspond to the definition of 'creative nodes' resembling the structural holes creative persons fill in social networks [153]. Creative node positions display a discrete dynamic behavior [32,33] and were termed as independent dynamic segments [69]. Importantly, since there is a whole family of network centrality measures [154], and network dynamics should also be taken into account, the identification of key functional residues will certainly be enriched and improved further in the future. The above approaches can also be used in the identification of drug binding sites as we will discuss in Section 3.D. in more detail. As an example of these utilizations, protein structure network position-based additional scores significantly improved the rigid-body docking algorithm of pyDock [152].

In the last few years automated servers and analysis methods have been established to convert Protein Data Bank 3D protein structure files into protein structure networks. The RING server (http://protein.bio.unipd.it/ring) [155] gives a set of physico-chemically validated amino acid contacts, and imports it to the widely used Cytoscape platform [Shannon *et al.*, 2003] enabling their further analysis. Using this server additional secondary structure-, solvent accessibility-, and experimental uncertainty-based factors may be combined with information on residue conservation, mutual information and residue-based energy scoring functions. Doncheva *et al.* [132,133] published a set of papers developing a specific assessment tool-kit for protein structure networks, RINalyzer (http://www.rinalyzer.de) linked to Cytoscape, and complemented with a protein structure determination module, RINerator (this latter stores the pre-determined protein structure networks of Protein Data Bank 3D protein structure files at http://rinalizer.de/rindata.php). The RINalyzer program has also been linked to NetworkAnalyzer allowing the examination of residue interaction networks derived from protein structures, and the comparison of protein structure networks.

Evolutionary conservation patterns of amino acids in related protein structures identified protein sectors [156]. Protein sectors are sparse networks of amino acids spanning a large segment of the protein. Protein sectors are collective systems operating rather independently from each other. Segments of protein sectors are linked to many surface sites distributed throughout the entire structure, are correlated with protein movements linked to enzyme catalysis and sector-connected surface sites are often places of allosteric regula-tion [157]. Sectors emerge as an 'intra-protein wiring' transmitting perturbations between important sites of substrate, ligand, protein, RNA, or DNA binding. A similar concept has been published by Jeon *et al.* [158], who determined that co-evolving amino acid pairs are often clustered in flexible protein regions.

At the fastest level protein dynamics is usually described by the ensemble of atomic vibrations. Conformational changes occur at a slower dynamical level. While current methods of protein structure networks usually use them for the structural description of proteins, the statistical mechanics approaches of Gaussian network models, elastic network models and normal-mode analysis are mostly used to assess protein dynamics. In elastic network models a harmonic potential is used to describe pairwise interactions between αC atoms only, or among all atoms, forming a spring network [23,28-37]. An increasing number of publications use correlation networks of ensemble-based models of the native state, residue position fluctuations, or covariance of NMR chemical shift changes [159-161]. These models helped the localization of hot-spots for ligand binding, and the identification of information pathways in proteins.

Conformational networks describe all possible conformations of a given protein, where nodes are the individual conformations the protein may adopt, and edges represent the possible transitions between these conformations [23,37,69,162]. The possible interrelationships between protein structure networks and conformational networks have not been explored so far and represent a truly exciting area of further studies.

### 3.B. Network Representations of Conformational Changes

Conformational changes seldom 'just happen'. Most of large-scale conformational changes have to be provoked. Provoked by the binding of a substrate, ligand or drug, or another macromolecule be it a protein, RNA, DNA or membrane, or by a change in the physico-chemical environment, such as a change in pH, molecular crowding, hydrophobicity of the surrounding medium, change in hydrogen bond structure, etc. The above considerations also predict that, if conformational change happens, it is usually not a smooth and straightforward process. Indeed, a rather long time ago 'protein quakes' have been described, i.e. a discontinuity of the conformational change was discovered [163]. A conformational change is often an avalanche of incremental, small changes in chemical bond configurations. If the first of these bond-rearrangements triggers the second, which triggers the third, we may finally arrive at a highly concerted, highly cooperative, but at the same time, at the level of the individual molecules, highly unpredictable process. The protein quake leads to a conformational change if it was able to cover the whole length of the trajectory leading to the appreciable change, and was not stalled somewhere in the middle, or swung back to the original position again.

In the last years many databases on conformational changes, like DynDom (http://fizz.cmp.uea.ac.uk/dyndom/enzymeList.do), ASD (http://mdl.shsmu.edu.cn/ASD/) have been assembled [164,165]. DynaSIN (http://archive.gersteinlab.org/proj/DynaSIN/) allows the investigator to assess the effects of conformational changes on the protein-protein interaction network [166].

Binding of the conformation-provoking agent requires a gross reorganization of interactions (including desolvation), which is a source of frustration and conflict in need of efficient mediation to accomplish binding [167-171]. Mediation of conflict may be provided by

1)  key residues of protein structure, which may be positioned at inter-modular boundaries, [170] and which we called independent dynamic segments [32,33,69,131,153];
2)  transient bonds (e.g. transient, non-native hydrogen bridges) [170,172,173];
3)  water [162,170,171] and by
4)  molecular chaperones [162].

It is important to note that all the above forms of mediation are rather transient. Independent dynamic segments [32,33,69,131] may lose their independence as binding proceeds. Before binding, by definition, independent dynamic segments have separate dynamics from the rest of the protein. Binding-induced stabilization may coalesce the motion of these key protein segments with neighboring domains. (In fact, this may contribute to the binding-induced stabilization of the novel structure.) Water molecules are expelled by the binding-induced, gradual desolvation [167-171]. Chaperones release their targets aiding the accomplishment of protein complex formation. In all cases the decrease of mediation is gradual, and occurs in concert with the decreasing need for mediation of the conflicts. All the above events are occurring at a rather large scale if the binding partner is a macromolecule, like a protein. However, most of them play a role, if the binding partner is a small molecule, like a substrate, ligand or drug, or the conformational change was triggered by even smaller individual triggers posed by the change in the environment.

As binding proceeds not only the conformation of the protein is changing (accompanied by the changing position of the protein on its energy landscape) but the approaching binding partner also changes the shape of the energy landscape itself. As the protein is approached by its binding partner, electrostatic and water-mediated hydrogen-bonding signals emerge, and they increasingly change the protein's environment altering its energy landscape [69,162,174]. The encounter of the protein with its binding partner involves a large number of conditional steps, where the next step of the encounter by the protein depends on a preceding change of its partner and vice versa. These concerted, conditional steps are not that well developed if the binding partner is a small and rigid molecule, like a substrate, ligand or drug, since the rigid binding partner does not have a lot of degrees of freedom to adapt. However, if the binding partner itself is also a protein or another macromolecule (RNA, DNA), the process becomes similar to 'wooing', and may be called an 'interdependent protein dance' [69,162].

The mutually conditional, stepwise encounter and selection steps can be modeled as a series of common social dilemma games. In the hawk-dove game model the rigid binding partner is the hawk, while the flexible binding partner is the dove, the payoff is the decrease of the free energy of the unbound state. The Nash equilibrium in the case of a hawk/dove encounter corresponds to an induced fit type scenario, where the conformational change of the dove is much larger than that of the hawk. This is a rather common case if a protein binds to a ligand. Here the protein is the dove, and the ligand is the hawk. The ligand (hawk) is the winner as compared to a flexible protein (a dove). For the rigid ligand the enthalpy gain of binding is not accompanied by an entropy cost as opposed to the more flexible protein, since the latter loses several degrees of freedom during binding. The ultimatum game provides another representation of the same scenario resembling the induced-fit mechanism. Here the two binding partners want to share the free energy decrease as the resource. The first, rigid partner proposes how to divide the sum between the two partners, and the second, flexible partner can either accept or reject this proposal [69,162, 174-176].

We note that the duality of the start state and the end state inferred by the concept of conformational change is not unique to protein structures, but can be generalized to all complex systems. A 'conformational change' starting from a rather integrated structure under normal conditions and changing to a rather modular structure upon stress was recently observed in the yeast interactome. Changes in modular structure were shown and proposed to be a general mechanism of adaptive processes. In a neural network model it was shown that a modular structure of the neural network gives a faster convergence of the neural network behavior to quasi-stable states (attractors). This corresponds to the above assumption that a certain level of modularity accelerates the adaptive changes. Bimodal behavior is characteristic to many types of networks including protein-protein interaction networks, metabolic networks, social networks and last, but not least protein structure networks [131,149,177-179]. The emergence of preferred (often bimodal) final states has been demonstrated by several studies calling the preferred states attractors in the state space (corresponding to a multidimensional energy landscape in case of molecular or cell rearrangements). Cancer was proposed to be a divergence from the normal trajectories representing the actual changes of cells, which had a smaller barrier due to mutations or epigenetic changes allowing malignant transformation. It turned out that between 80 and 200 changes in gene transcription must contribute to complete this process. Approach of the attractors representing various stable states of the cells was proposed to have multiple trajectories. Trajectories can be deconvoluted to gene-encoded regular trajectory segments and to the stochastic variations of these regular trajectory segments caused by the actual, unique environmental factors [180-182]. For the transition from one attractor to the other (like during the differentiation of hematopoietic progenitor cells into the eyrthroid or neurophil lineages) networks have to acquire a transient flexibility to allow the transition to occur [183]. Please note that many of the above considerations in other, mostly cellular systems hold to the conformational changes of proteins. The multiple pathways converging at the other attractor, the transiently increased flexibility, the importance of modular structure have already been uncovered in examining the mechanism of conformational changes as we will show below, or provide exciting research areas for future studies.

The mechanism of conformational changes attained a lot of attention in the last century. In these studies quite often allosteric interactions have been assessed. The reason of this is that in allosteric interactions the final effect of the allosteric ligand to enzyme activity could be assessed quite readily. In recent years precise analytical methods for the dynamics of protein structures became available, which allowed the analysis of conformational changes which were not linked to changes in an enzyme activity [69-72,184-186].

The first step of binding is quite often the formation of transient encounter complexes. These complexes are mostly stabilized by electrostatic forces, have a small, planar contact area in the range of only a few $Å^2$, and cover a larger segment (e.g. 15%) of the total surface area around the binding site as revealed by paramagnetic relaxation enhancement experiments [187,188].

In the following step 'anchor residues', i.e. residues stabilizing the transient encounter complex formed in the first step, often play an important role. Anchor residues have a similar position as in the bound state, and have a large surface area in the range of 100 $Å^2$. Binding to anchor residues is followed by the completion of binding involving several 'latch residues' giving a further stabilization of the complex [189].

Anchor and latch residues (positioned often very close to the binding site) are by far not the only protein segments involved in binding partner-induced conformational changes. Hinges (located often at distant positions from the binding site) were proposed to play a decisive role in these conformational transitions [172]. Critical nodes between communities of amino acid networks were also shown to play a key role in the reorganization of the protein structure [23,145,153,190]. Recently elastic network analysis uncovered the existence of specific protein segments having 'discrete breather' behavior. These segments accumulate and dynamically exchange a large portion of overall energy, and are often located very close to the binding or catalytic sites [32,33]. We called these one to a few amino acid long parts of protein 'independent dynamic segments' to emphasize their distinct behavior [69]. Hinges, critical nodes and independent dynamic segments significantly overlap, since independent dynamic segments are located in stiff, hinge-type regions of proteins, and both may often be positioned at inter-modular boundaries [23,32,33,69,145,153,172,191]. A key role of specific regions of co-evolving amino acids called 'protein sectors' in the transmission of conformational changes was also suggested [157]. Currently it is not known to what extent hinges, critical nodes, independent dynamic segments and protein sectors define the same protein segments.

As we described above, the inventory of individual protein segments involved in the orchestration of conformational changes has been grossly expanded in the last few years. However, the sequence of dynamic events, how they interact, and how they actually cause the change, is still not completely understood. A series of conformational distortions has been proposed more than ten years ago to mediate the conformational change [160]. Pathways of conformational change propagation (mostly in allosteric changes where both the trigger site and the effector site can be easily identified) have been uncovered using protein structure networks, Gaussian network models, Markov random walks, correlations in residue fluctuations, or co-variance of NMR chemical shifts. Propagation pathways were small-worlds providing rather short routes between the trigger and the effector sites and often involved multiple trajectories converging at inter-domain boundaries [159,161,190,192-199].

Current molecular mechanisms of conformational changes include the propagation of a 'frustration front' [37], where frustration means that certain residues have accumulated a tension, which prompts their unusually large dislocation, once the allosteric ligand triggered the release of this tension. A recent study [200] on 51 pairs of proteins in active and inactive state using the FIRST algorithm [31] found that rigid paths connect the effector and catalytic sites in 69% of the data set. An increased rigidity in active state relative to the inactive state has also been observed. However, 31% of the studied structures failed to follow this trend [200]. The uncertainties prompt further studies of this exciting field.

### 3.C. Network Analysis of Protonation Changes

Protonation changes have seldom been studied directly using protein structure networks. Most protonation-related studies dealt with hydrogen bonding networks, where the word "network" is used in a different context, like elsewhere in this review. Hydrogen bonding 'networks' consist of a rather few nodes, seldom reaching ten. These very tiny networks still play a paramount role in the mechanism of active centers of several enzymes and in many water-mediated conformational changes, such as those of the activation of family A G-protein-coupled heptahelical receptors [201] and of mineralocorticoid receptors [202], as well as in the Ca-decrease-dependent activation of the tobacco mosaic virus protein [203]. The midpoint reduction potential of cytochrome c is also controlled by a hydrogen bond network [204]. Interestingly, a water molecule plays an important role in the stabilization of the interaction of HIV-1 protease with its inhibitors acting as a collision buffer, and helping the re-formation of broken hydrogen bonds [205]. The activity of subtilisin and chymotrypsin inhibitor 2 also depends on an extensive hydrogen bonding network [206,207]. A particularly nice example of the involvement of protonation-related networks is the activation of native rhodopsin, which proceeds via the protonation of the retinal Schiff-base, where coupling of several protonation switches has been shown [208-210]. Re-design of hydrogen bond network may grossly extend the specificity between the proper substrate (drug) and its binding protein [211].

Protonation changes of the bacterial Resistance-Nodulation-Division (RND) family drug/proton antiporters led to gross conformational rearrangement mainly at charged residues in the transmembrane domain helping the extrusion of the drug bound to the periplasmic domain of the RND transporter [212]. With the RND proton antiporter protonation-driven conformational changes are more than expected. Whether such a mechanism might exist in ABC transporters is a question of interesting future studies.

### 3.D. Pharmacological Applications of Protein Structure Networks

Protein structure networks have been involved in many steps of drug design from the determination of drug binding sites to the identification of novel drug targets. In this review we give only a brief summary of the most important applications.

The first step to identify drug binding sites is to find appropriate cavities on protein surfaces. Though excellent programs exist to serve this purpose such as the SCREEN [213], which also takes into account physico-chemical, structural and geometric properties of the cavity, protein structure networks have seldom been used yet for this purpose.

Quantitative structure-activity relationship (QSAR) methods have an unparalleled importance in the identification of optimized drug binding. Protonation states may be incorporated into usual 3D receptor QSAR methods as a $4^{th}$ dimension (4D QSAR) leading to more accurate results [214,215]. As an additional source, PDTD (http://www.dddc.ac.cn/pdtd/) is a web-accessible protein database for drug target identification, docking a small molecule to a repository of proteins with known 3D structure [216].

Binding of a large number of drugs, or drug-like molecules to proteins can be studied by wet lab or *in silico* experiments. An educating example for the former is the study of Clemons *et al.* [217], where a diverse collection of small molecules was bound to 100 diverse proteins. Increasing the content of $sp^3$-hybridized and thus stereogenic atoms improved binding selectivity and frequency. An *in silico* study showed that druggable sites comprised a cluster of binding hot-spots having a concave topology with a pattern of hydrophobic and polar residues. Importantly, druggable sites had a conformational adaptivity to allow the hot-spots to expand, and accommodate a ligand of drug-like dimensions [218].

Network properties of drug-binding sites are rather similar to those of active centers and of the residues involved in the propagation of conformational changes detailed in Section 3.B. Residues of heme-binding regions were found to be generally associated with hubs, high closeness and betweenness centralities together with a low clustering coefficient [151]. The interesting study of Asses *et al.* [219] extended the known properties of the catalytic site of the Met-kinase, an interesting target for anticancer drug design. The ATP binding site of this kinase displays an unusually high plasticity, which may allow its selective targeting. Previously unknown conformations of the site were modeled and serve the design of novel Met-selective drug candidates.

Recently a number of protein/ligand or protein/drug databases were established. CREDO (http://www.cryst.bioc.cam.ac.uk/credo) lists protein/ligand interactions, representing contacts as structural interaction fingerprints [220]. DrugBank (http://www.drugbank.ca) is a dataset of the interactions of more than 4,100 drugs and more than 14,000 proteins [221]. STITCH 3 (http://stitch.embl.de) offers an aggregated database of the interactions over 300,000 chemicals and 2.6 million proteins [222]. Dr. PIAS (http://www.drpias.net) assesses whether a protein-protein interaction is druggable as a target for small chemical ligands [223]. Besides the assessment of the binding profile of new drug candidates, the use of protein/drug databases may lead to the discovery of unexpected novel targets of known drugs, as was the case when anti-Alzheimer drugs have been assessed [224].

Designing drugs for targets with similar active sites is challenging due to low specificity of the molecules on the chosen target. Construction and analysis of binding-site similarity networks can be helpful to identify proteins whose active sites are different enough to be targeted selectively. Using 491 human protein kinase sequences, Huang *et al.* [225] constructed similarity networks of kinase ATP binding sites. The selective inhibition of kinases that are in small clusters of the network having only few links to other nodes, is more likely than of those having a large number of links in large clusters, hence, being similar to many other kinases. This approach to selective inhibition is consistent with recent experimental results targeting tyrosine kinase EphB4 that belongs to a small, highly linked subfamily of kinases with few links to other kinases [225].

Allosteric drugs (binding to allosteric effector sites) are considered to be better than orthosteric drugs (binding to active centers) due to three reasons. First and foremost: allosteric drugs often have less side effects than orthosteric drugs due to the larger variability of allosteric binding sites than that of active centers. Second: they allow the modulation of drug therapy effects in a tunable fashion. Third: they mostly act only when the endogenous ligand is bound, i.e. exactly at the time when the cell needs their action. They are non-competitive with the endogenous ligand therefore their dosage can be low [226,227]. Allosteric drugs may not need to bind to '*in vivo*' allosteric sites, but may also bind to allosteric sites which are not used by natural ligands. The intra-protein propagation pathway of the allosteric effect becomes increasingly known as we summarized in Section 3.B. and as it is described in the review of De Benedetti and Fanelli [195]. Using this knowledge novel allosteric drug binding sites might be found on several proteins. As a potential novel source for such studies, non-synonymous single nucleotide polymorphisms (nsSNPs, SAPs) were found to be associated with diseases if the changed amino acid was a hub, or had a high centrality in the protein structure network. The neighboring residues around the SAP site quite strongly determined its association with disease [228]. Disease-associated SAPs may often be a part of the propagation pathways of allosteric effects. Their surface-associated subset may be an interesting dataset to assess the existence of novel allosteric drug binding sites.

The concept of allosteric drugs can be broadened to allo-network drugs, whose effects can propagate across several proteins, to enhance or inhibit specific interactions along a pathway. Drug action propagates from the original binding site to protein-protein interaction network (interactome) neighbors in a non-isospheric (i.e. not uniform to all distances), highly specific manner. Binding sites of promising allo-network drug targets are not parts of 'high-intensity' intracellular pathways, but are connected to them. Thus allo-network drugs can achieve specific, limited changes at the systems level achieving fewer side effects and lower toxicity than conventional drugs. Allosteric effects can be considered at two levels: (1) stepwise events restricted to the neighbors or network module of the originally affected protein; (2) propagation via large cellular assemblies over large distances (i.e. hundreds or even thousands of Angstroms) [227].

For the identification of allo-network drugs first the interactome has to be extended to atomic level resolution. For this, docking of 3D protein structures and the consequent connection of their protein structure networks are needed. A successful candidate for the inter-protein allosteric pathways involved in allo-network drug action disturbs the network perturbations specific to a disease state of the cell at a site distant from the original drug-binding site. Perturbation analysis [174,229] applied to the atomic level resolution of the interactome in combination with disease specific protein expression patterns may help the identification of such allo-network drug targets. We may use a number of centrality measures including perturbation-based, or game-theoretical assumptions to find the level of importance of various proteins and pathways in the cellular network [154,229]. By the analysis of directed networks, such as signaling or metabolic networks, we may construct a tree-like hierarchy [230] to assess the importance of various nodes, or we may find driver nodes controlling the network by the application of a recently published method [231]. A general strategy for the identification of allosteric sites may involve finding large correlated motions between binding sites. This can reveal which residue-residue correlated motions change upon ligand binding, and thus can suggest new allosteric sites [232] even in integrated protein structure net-works of mega-complexes. Reverse engineering methods [233] allow us to discriminate between 'high-intensity' and 'low-intensity' communication pathways of cellular networks, and thus may provide a larger safety margin for allo-network drugs. In this way a large number of novel and specific action sites can now be proposed as potential targets of allo-network drugs.

A specific extension of orthosteric, allosteric and allo-network drugs are multi-target drugs, which act on multiple target sites. Multi-target drugs multiply the number of pharmacologically relevant target molecules by introducing a set of indirect, network-dependent effects. Many conventional drugs, like aspirin are multi-target drugs. Combination therapies also use the effect of multi-target action. Simultaneous binding to multiple targets requires mutual compromise in binding accuracy and results in a lower affinity than the precise targeting of a single protein. Our earlier studies showed that multiple actions with lower affinity (50% efficient) may result in similar effects at the whole network level than a 100% efficient targeting of the most important network hub, even if only 3 to 5 proteins are targeted [128]. Designing multi-target drugs is a challenging but not a formidable task to solve. Moreover, the low-affinity binding of multi-target drugs eases the constraints of druggability and significantly increases the size of the druggable proteome [129,130].

How to find proper multi-target sets? All the above considerations for allo-network drugs can be extended allowing multi-target design. Multiple perturbations may be assessed [174,229], centrality, hierarchy, controllability and co-variance measures [154,231, 232] can be calculated in a multi-nodal fashion. Xie *et al.* [234] in an attempt to discover drugs that are able to inhibit multiple targets constructed similarity networks of protein binding sites. The authors generated a network of 232 non-redundant *M. tuberculosis* proteins using the sequence-order-independent profile-profile alignment (SOIPPA) algorithm, and identified the 18 mostly connected proteins. Identifying clusters of proteins with similar binding sites could be another starting point for designing multi-target drugs.

Several network considerations are seemingly unrelated to the problem of multi-target drug design, but could still be used quite efficiently to solve this complex problem. We just give two recently published examples here. Morris and Barthelemy [235] examined the transport between multiple sources and sinks in a coupled network, which is a system where two interdependent networks are tightly linked together. They found that the number of sinks determines the optimal routes from sources, which is the minimum shortest path if the number of sinks are low, but congestion phenomena occurs, and necessitates a more complex treatment when the number of sinks increases. These considerations may well be used when designing multi-target drugs. An interesting paper gave an algorithm to describe the controllability of leader-follower networks of multi-agents systems [236]. Though the systems they covered included robotic networks, flocking birds, sensor coverage and others, the method might also be useful in multi-target drug design.

## 4. CONCLUSION AND PERSPECTIVES

In this concluding Section first we give a few examples how the network approach might be used in the assessment and modulation of ABC transporter functions. From the examples above it is clear that the proper folding, expression, and function of ABC transporters is maintained by and is highly sensitive to dynamics of the polypeptide chain. The different dynamics of various parts of a protein exist not as an individual units but exhibit highly coupled motions. The analysis of the coupling of these dynamic units employing a network approach may help to understand the mechanism of function and to identify drug targets.

For example, it became evident that targeting ΔF508 CFTR at a single location to improve either its kinetic or thermodynamic properties rescued the mutant channel only inefficiently [127]. Based on common sense, all properties must be corrected to fully restore the functional expression of the mutant protein. However, principles of network theory suggest that partial modulation of multiple dynamic nodes in the protein will have a greater effect on the network (structural and dynamic units of the protein) than the complete modulation of one of the nodes (kinetics or thermodynamics) [128,129].

The rescue of ΔY490 MDR1 by a hydrophobic substrate suggests that the dynamics of the transmembrane regions is strongly coupled to the nucleotide binding domains [124]. This also indicates that the design of small molecules acting in the transmembrane regions is rational to correct the folding and dynamics of intracellular domains without the need for the corrector molecule to enter the cytoplasm. In order to avoid designing substrate-like small molecules that would interfere with the transport of the physiological substrate of transporters, modulation at allosteric sites would be desirable. A network approach can be applied to identify dynamic units which are also allosteric sites and connections between them. Such a network map would enable one to select the appropriate unit (node) whose dynamics is to be modulated in order to functionally rescue the protein. A similar network approach can be used to shift the dynamic ensemble of a mutant with altered substrate specificity to the ensemble characteristic of the wild type protein to avoid side effects of drugs tested in individuals expressing the wild type transporters.

In summary, in this review we detailed the importance of synthesis of structural, dynamic, computational and experimental information on ABC proteins to design and improve strategies modulating their functional expression. We showed how the dynamics of different parts of ABC proteins are coupled and the importance of studying intramolecular interactions using network approach. Now we will highlight the major points where we predict progress in this pharmacologically exciting expanding field:

- detailed comparison of various network descriptions, such as protein structure networks, protein sectors, Gaussian network models, elastic network models, normal mode analysis, networks based on the correlations of residue fluctuations and conformational networks will bring a much detailed understanding of both protein structure and function;
- detailed analysis of conformational change space in terms of trajectories and attractors will greatly benefit from the network approach in the future;
- refined modular structure of networks related to protein structure and function will certainly give a novel understanding of structure/function relationships;
- the identification of key functional residues will benefit from the recent introduction of novel structural and dynamic network centrality measures;
- the mechanism of conformational changes is far from being understood, here network studies may bring novel solutions and understanding;
- protonation changes have only been seldom covered by network studies so far;
- several pharmacological applications remain uncovered, here network approach may lead to the discovery of novel drug binding sites and drug targets.

All the above considerations will bring new areas to uncover ABC transporter action and for its pharmacological modifications. Comparison of various conformations of ABC proteins using global network invariants will certainly help the understanding of ABC transporter function. Comparison of nodes in different conformations having a specific network position, such as those described by Guimera *et al.* in their paper on network cartography [237] will extend our knowledge on structure/function relationships. Studies on the modular structure [149,154] of ABC transporter protein structure networks (as illustration see Fig. **3**) may reveal functionally important, inter-modular residues, or residue-set, acting in concert. Perturbation analysis [174,229] of ABC transporter protein structure networks may highlight protein segments which have a key role in the transmission of conformational changes. Game theoretical assessment [69,162,174] of the cooperation of ABC transporter protein structure networks and their modules may highlight novel functional units. Inter-modular residues (or residue-sets) and modular centers of ABC transporter protein structure networks determining intra-protein information transmission, as well as their cooperative units may all be proven highly important in pharmacological targeting of ABC transporters. The assessment of multi-target drug action [129] or allo-network drugs [227] in the modification of ABC transporter action may also give novel pharmacological approaches in this important field.

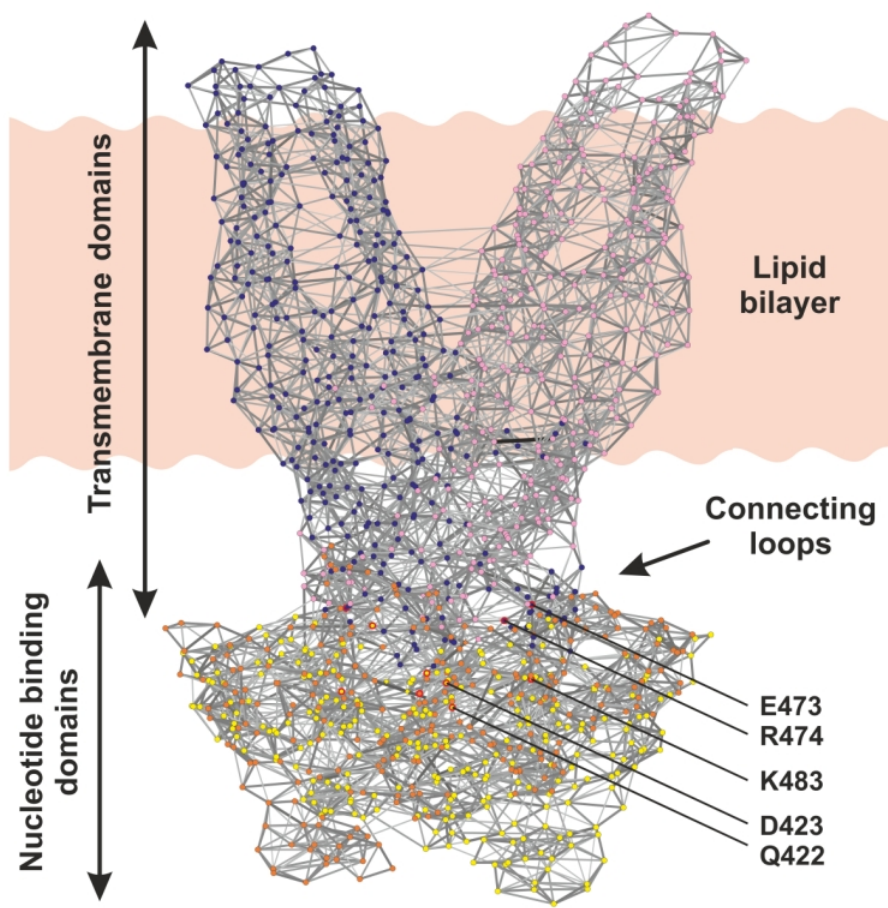

**Fig. (3).** Modules of ABC transporter protein structure network. The protein structure network of the ABC transporter Sav1866 (PDB ID: 2HYD) was assembled using the HBPlus program [240] by converting the Cartesian coordinates of the 3D image to distances of amino acid pairs, and keeping all non-covalently bonded contacts within a distance of 0.4 nm and applying weights inversely representing the distance between amino acid residues. Modular analysis was performed by the Cytoscape plug-in of the ModuLand method [149]. On the figure modules of the Sav1866 protein structure network at the fourth hierarchical layer are shown (in each hierarchical layer modules of the layer below became meta-nodes of the upper layer and overlaps between modules of the layer below became edge-weights of the upper layer). Modules of the 4$^{th}$ hierarchical layer are projected back to the original protein structure network consisting of Sav1866 amino acids as nodes to allow a better comparison and are labeled with different colors. The network was visualized using the Pajek program [241]. Amino acids participating in the electrostatic interactions at the transmembrane domain and nucleotide binding domain interface [239] are highlighted with red contours and are labeled with arrows, while (small) diamonds label water molecules. Note that both the amino acids participating in the inter-domain electrostatic interactions and the connecting loops are situated at the overlap of major modules of the protein structure network. (The color version of the figure is available in the electronic copy of the article).

We are at the very beginning of the understanding of ABC protein dynamics, as well as of the elucidation of their mechanism of action. However, from the studies summarized in our paper, it is already clear that the combination of computational and experimental tools to study the effect of physiologically important alterations in dynamics of ABC proteins will provide a very promising research area in the coming years.

**CONFLICT OF INTEREST**

The authors confirm that this article content has no conflicts of interest.

**ACKNOWLEDGEMENTS**

Authors would like to thank Eszter Hazai (Virtua Drug Co., Hungary) for her help in construction protein structure networks, and members of the LINK-group (www.linkgroup.hu) for helpful suggestions. We thank Balazs Sarkadi (Membrane Research Group, Hungarian Academy of Sciences, Budapest, Hungary) for critically reading the manuscript. Work in the authors' laboratory was supported by research grants from the Hungarian National Science Foundation (OTKA-K83314, MB08C-80039), from the EU (TÁMOP-4.2.2/B-10/1-2010-0013, FP7-IRG 239270) and by a residence at the Rockefeller Foundation Bellagio Center (PC). T.H. is a Bolyai Fellow of the Hungarian Academy of Sciences.

**ABBREVIATIONS**

| | | |
|---|---|---|
| ABC | = | ATP binding cassette |
| ADP | = | Adenosine diphosphate |
| Arg | = | Arginine |
| ATP | = | Adenosine triphosphate |
| CFTR | = | Cystic fibrosis transmembrane conductance regulator |
| EPR | = | Electron paramagnetic resonance |
| ER | = | Endoplasmatic reticulum |
| Gly | = | Glycine |
| MDR | = | Multidrug resistance |
| MDR1 | = | Multidrug-resistance protein 1 |
| MDR1A | = | Multidrug-resistance protein 1A |
| MRP1 | = | Multidrug-resistance associated protein 1 |
| NBD | = | Nucleotide binding domain |
| NMR | = | Nuclear magnetic resonance |
| nsSNPs | = | Non-synonymous single nucleotide polymorphisms (SAPs) |
| Phe | = | Phenylalanine |
| $P_i$ | = | Inorganic phosphate |
| PKA | = | Protein kinase A |
| PM | = | Plasmamembrane |
| QSAR | = | Quantitative structure-activity relationship |
| RI | = | Regulatory insertion |
| RIN | = | Residue interaction network |
| RND | = | Resistance-Nodulation-Division |
| SAPs | = | Non-synonymous single nucleotide polymorphisms (nsSNPs) |
| Thr | = | Threonine |
| TM | = | Transmembrane |
| TMD | = | Transmembrane domain |